\RequirePackage{lineno}
\documentclass[pdflatex,aps,prd,twocolumn,nofootinbib,showpacs,superscriptaddress]{revtex4-2}
\usepackage{amsmath,amssymb,color,hyperref}
\usepackage{graphicx}
\usepackage[latin1]{inputenc}
\usepackage[english]{babel}
\usepackage{booktabs, multirow, makecell}%
\usepackage{comment}
\usepackage{capt-of}
\usepackage{float}

\begin{document}
\title{Lee-Yang and Langer edge singularities from analytic continuation of scaling functions}

\author{Frithjof Karsch$^1$, 
Christian Schmidt$^1$ and Simran Singh}
\affiliation{Fakult\"at f\"ur Physik, Universit\"at Bielefeld, D-33615 Bielefeld,
Germany}

\begin{abstract}
We discuss the analytic continuation of scaling function in the 3-dimensional $Z(2)$, $O(2)$ and $O(4)$ universality classes using the Schofield 
representation of the magnetic equation of state. We show that a determination of the location of
Lee-Yang edge singularities and, in the case of 
$Z(2)$, also the Langer edge singularity yields
stable results. Results for the former are in good agreement with Functional Renormalization Group calculations.
We also present results
for the location of the Langer edge singularity in the $3$-$d$, $Z(2)$ universality class. We find that in 
terms of the complex scaling variable $z$ the distance of the Langer edge singularity to the critical point
agrees within errors with that of the 
Lee-Yang edge singularity. 
Furthermore the magnitude of the discontinuity along the Langer 
branch cut is an order of magnitude smaller than that along the Lee-Yang branch cut.
\begin{center}
\bf{\today}
\end{center}
\end{abstract}

\vspace{0.2in}

\newcommand \tb {\bar{t}}

\maketitle

\section{Introduction}

In the vicinity of a phase transition point the universal critical behavior
of thermodynamic observables is parameterized in terms of scaling functions.
In studies of the phase structure of
the theory of strong-interactions,
Quantum Chromodynamics (QCD), 
scaling functions in the 3-$d$,
$Z(2)$, $O(2)$ and $O(4)$ universality classes 
play an important role.
In particular, when studying QCD 
at non-zero temperature and with
non-vanishing chemical potentials that
are the external control parameters for the presence of non-vanishing  conserved charge densities, 
the $O(4)$ universality class controls
properties of QCD in the chiral limit 
while the $Z(2)$ universality
class controls the critical behavior
in the vicinity of the so-called critical
end-point that is expected to exist
for large values of the baryon chemical potential
 \cite{Gross:2022hyw,Philipsen:2021qji,Guenther:2022wcr}.
When analyzing the phase  structure of QCD
on discrete space-time lattices parts of the global symmetries are broken explicitly and smaller symmetry groups,
e.g. $O(2)$, start playing a role.

In QCD with real, non-vanishing chemical
potentials the direct application of numerical methods fails because the 
integrand of the path integral, defining
the QCD partition function,  becomes complex. 
One tries to circumvent this problem by
either performing numerical calculations with imaginary chemical potentials \cite{DElia:2002tig,deForcrand:2002hgr} or by using Taylor series expansions
\cite{Gavai:2001fr,Allton:2002zi,Bollweg:2022rps}, which are set-up 
at vanishing values of the chemical potential.
In both cases it is of importance to
understand the analytic structure of
the QCD partition function and get control over the location of singularities in the complex plane 
that hamper analytic continuations
on the one hand and limit the 
radius of convergence of Taylor series
on the other hand. This has led to recent
interest in the analysis of analytic properties of the universal scaling 
functions \cite{PhysRev.87.404,PhysRev.87.410} in universality classes that 
are of relevance for studies of the QCD
phase diagram 
\cite{An:2016lni,An:2017brc,Mukherjee:2019eou,Mukherjee:2021tyg,Basar:2021hdf,Basar:2021gyi,Dimopoulos:2021vrk,Schmidt:2022ogw,Clarke:2023noy}.

Scaling functions,
which describe the singular, universal
part of thermodynamic observables,
e.g. the order parameter ($M$) and 
its derivatives with respect to temperature
$T$ or external field parameter $H$,
are commonly parameterized as function of the real scaling variable 
$z=z_0 H^{-1/\beta\delta} (T-T_c)/T_c$. 

Prominent features of the analytically continued scaling functions
in the complex $z$ plane  are the
occurrence of edge singularities that mark the end-point of branch cuts.
In the $Z(2)$ universality class these
are the Lee-Yang edge singularity \cite{PhysRev.87.404,PhysRev.87.410} 
and the so-called Langer edge singularity \cite{Langer:1967ax}.
In fact, these singularities are expected
to be closest to the origin and
thus limit the radius of convergence 
of Taylor series.
Both singularities have been discussed extensively \cite{An:2016lni,An:2017brc}.
While the phase of the associated cuts in
the plane of complex valued scaling variable $z$ is rigorously known to be $\phi_{LY}=\pi/2\beta\delta$ for the 
Lee-Yang cut and $\phi_{Lan}=\pi(1-1/\beta\delta)$
for the Langer cut, the absolute 
values of the Lee-Yang edge singularity has only been determined recently in Functional
Renormalization Group (FRG) calculations \cite{Connelly:2020gwa,Johnson:2022cqv}.

Rather than using $(T,H)$ for the parameterization of thermodynamical
observables another set of parameters, $(\theta,R)$,
has been introduced by Schofield
\cite{PhysRevLett.22.606,PhysRevLett.23.1098}.
This provides a convenient parameterization of the the magnetic
equation of state \cite{Widom,PhysRev.158.176}, making its
dependence on critical exponents explicit.
Approximations for the 
unknown function $h(\theta)$ that 
enters this parameterization have been determined
in analytic as well as numerical 
calculations \cite{Engels:1999wf,Karsch:2023pga}.
In cases where the analytic form of $h(\theta)$ is exactly known 
the Schofield parameterization provides straightforwardly an analytic continuation of the scaling functions 
into the complex $z$-plane. This is,
for instance, the case in the mean-field approximation (MFA) as well
as the $N= \infty$ limit
of the 3-$d$, $O(N)$ models. We will
discuss this here. The main goal of this work,
however, is to examine the analytic continuation
of scaling functions using the Schofield 
parameterization with only approximately known
functions $h(\theta)$ as it is currently the case
in the $Z(2)$ and $O(N)$ universality classes. 
We show that a determination of the location of
Lee-Yang edge singularities in these universality classes provides results in good agreement with the FRG analysis
\cite{Johnson:2022cqv}. Furthermore,
we provide first results for the location of the Langer edge singularity in the analytically
continued scaling functions of the 3-$d$, $Z(2)$ universality class.
We also show that the Langer and
Lee-Yang cuts can be identified, although details of the expected 
analytic structure in the vicinity
of these edge singularities 
\cite{Langer:1967ax,PhysRev.87.404,PhysRev.87.410,PhysRevLett.40.1610,Fonseca:2001dc,An:2017brc}
can not be reproduced 
when using only a truncated series expansion for the function $h(\theta)$.

This paper is organized as follows. 
In Sec.\ref{sec:Schofield} we summarize
basic definitions of scaling functions
in 3-$d$ universality classes and introduce their representation using
the Schofield parameterization of the magnetic equation of state. Sec.~\ref{sec:LPM} is devoted to a discussion of scaling functions in MFA, the large-$N$
limit of $O(N)$ models, and the ${\cal O}(\epsilon^2)$ approximation in 
the $Z(2)$ universality class, where
the generalized linear parametric model (LPM) provides exact results in the 
Schofield parameterization. In Sec.~\ref{sec:beyond} we discuss our 
results for Lee-Yang edge and Langer edge singularities obtained from the 
analytic continuation of the Schofield
representation of scaling functions. 
Finally we give
an outlook and conclusions in Sec.~\ref{sec:conclusion}. Some 
details on the determination of the 
parameterization of scaling functions in the $Z(2)$ universality class are given in Appendix~\ref{app:Z2}.

\section{Scaling functions in the Schofield
parameterization}
\label{sec:Schofield}
\subsection{Universal scaling functions}
In the vicinity of a critical point the free energy density, $f(T,H)$, 
of a thermodynamic 
system contains a non-analytic component,
the so-called singular part, $f_f(z)$,
which is given in terms of a particular combination of the external control
parameters, temperature ($T$) and symmetry breaking field ($H$)\footnote{We consider here only the leading singular behavior and suppress universal, singular contributions from well-known, corrections-to-scaling
(see for instance: \cite{Hasenbusch:1999mw}).},
\begin{equation}
    \frac{f(T,H)}{T}\equiv -\frac{1}{V}\ln Z(T,H)
    = H_0 h^{1+1/\delta} f_f(z) +reg. \;\; ,
    \label{free}
\end{equation}
with 
\begin{equation}
    z= \frac{t}{h^{1/\beta\delta}}
    \;\; ,\;\; h\equiv H/H_0\; ,\; t=t_0^{-1} \frac{T-T_c}{T_c} \; .
    \label{z-h}
\end{equation}
Here the critical temperature $T_c$ as 
well as the scale parameters $H_0$, $t_0$
are non-universal parameters, whereas the critical exponents $\beta$, $\delta$ 
are universal and characteristic for different universality classes. This also is the case for the entire scaling function $f_f(z)$.

In Eq.~\ref{free}  "reg." denotes additional contributions from regular terms that can be ignored
close to the critical point. 
Thermodynamic observables  and their universal scaling behavior can then be described in terms of scaling functions that can be derived from $f_f(z)$. We will consider here 
the scaling behavior of the order parameter,
\begin{equation}
    M \;=\; - \frac{\partial f}{\partial H} \equiv h^{1/\delta} f_G(z)\; , 
    \label{M-z}
\end{equation}
and the magnetic ($\chi_h$) and 
thermal ($\chi_t$) susceptibilities,
\begin{eqnarray}
\chi_h &=& \frac{\partial M}{\partial H} = \frac{h^{1/\delta -1}}{H_0}
 f_{\chi}(z) \; , \nonumber \\
 \chi_t &=& - T^2 \frac{\partial M}{\partial T} =  -\frac{T^2}{t_0 T_c} h^{(\beta-1)/\beta\delta} f'_G(z)
  \; . 
\end{eqnarray}
The scaling functions $f_G(z)$ and 
$f_\chi(z)$ are related to $f_f(z)$,
\begin{eqnarray}
f_G(z) &=& -\left(1+\frac{1}{\delta}\right) f_f(z)
+\frac{z}{\beta\delta}f'_f(z) \; ,
\nonumber \\
f_{\chi}(z) &=& \frac{1}{\delta} \left( f_G(z) - \frac{z}{\beta} f'_G (z)
\right) \; .
\label{fchi}
\end{eqnarray}
These scaling functions have
been determined in the entire
range of real $z$ values using
asymptotic expansions for 
large $|z|$, Taylor series at
small $z$, and Monte Carlo
simulations for small and intermediate regions that are either matched to the asymptotic behavior \cite{Engels:1999wf,Engels:2000xw,Engels:2002fi} or have
been described in terms
of the Schofield
parameterization of the magnetic equation of state
\cite{Engels:2002fi,Karsch:2023pga}.
We will introduce the 
Schofield parameterization
in the following subsection.

\subsection{Schofield parameterization of scaling functions}

The scaling functions in the 3-$d$, $O(N)$
universality classes, including the $Z(2)$ universality class corresponding to $N=1$, as well as the $N= \infty$ limit, have been determined by using $\epsilon$-expansions and
field theoretic methods in 3-$d$ \cite{Guida:1996ep,Guida:1998bx,Campostrini:2000iw,Campostrini:2000si,Campostrini:2002ky} 
as well as numerically using Monte-Carlo simulations \cite{Engels:1999wf,Engels:2000xw,Engels:2002fi,Engels:2011km}. 
The scaling functions obtained in 
analytic calculations are commonly parameterized using the Schofield
parameterization \cite{PhysRevLett.22.606} of the Widom-Griffiths (W-G) \cite{Widom,PhysRev.158.176}
form of the magnetic equation of state,
\begin{eqnarray}
M &=& m_0 R^{\beta}\theta\;, \label{W-G-M}\\
t &=& R(1-\theta^2)\; , \label{W-G-t}\\
h &=& h_0 R^{\beta\delta} h(\theta)\; , \label{W-G-h}
\end{eqnarray}
where ($R$,\ $\theta$) denotes an alternative coordinate frame obtained from ($t,h$) as defined by Eq.~\ref{W-G-t} and \ref{W-G-h} \cite{PhysRevLett.22.606,PhysRevLett.23.1098}.

Using Eqs.~\ref{W-G-M}-\ref{W-G-h}
and comparing these with Eqs.~\ref{z-h} and \ref{M-z} 
one easily finds the 
relation between the scaling
functions expressed in terms
of $z$ and $\theta$, respectively,
\begin{eqnarray}
f_G(z)&\equiv&  f_G(\theta (z)) =
 \theta \left( \frac{h (\theta)}{h (1)}\right)^{-1/\delta}\; ,
\label{W-G-fg} \\
z(\theta) &=&  \frac{1-\theta^2}{ \theta_0^2-1 }\theta_0^{1/\beta}
        \left( \frac{h(\theta)}{h(1)}\right)^{-1/\beta\delta} . 
        \label{W-G-z}
\end{eqnarray}
Aside from an explicit dependence on 
the universal critical exponents, $(\beta, \delta)$,
scaling functions then depend on the
function $h(\theta)$, which usually is
represented by a polynomial in $\theta$,
containing only odd powers of $\theta$.
The function $h(\theta)$ is positive
in the interval $0<\theta<\theta_0$, 
with $\theta_0>1$ denoting its first positive real zero. 
The real $z$-axis is mapped onto the interval $0<\theta<\theta_0$. 
Obviously, $\theta=1$ corresponds
to $z=0$, $\theta=\theta_0$ corresponds
to $z=-\infty$, and $\theta=0$ corresponds to $z=\infty$.

The prefactors as well as the normalization with $h(1)$, appearing in Eqs.~\ref{W-G-fg} and \ref{W-G-z}, arise from
the scale parameters
$m_0$ and $h_0$ which can 
be chosen such that the conventional
normalizations for the scaling 
functions hold; {\it i.e.}
$f_G=1$ at $z=0$ and 
$f_G/(-z)^\beta\rightarrow 1$ for
$z\rightarrow -\infty$. This leads
to the choice \cite{Karsch:2023pga},
\begin{equation}
    m_0 = \frac{(\theta_0^2 - 1)^\beta}{\theta_0}\;, 
    \; h_0 = \frac{m_0^\delta}{h(1)}\; .
    \label{norm-m-h}
\end{equation}

The function $h(\theta)$ then provides a parameterization
of the magnetic equation of state. Using Eqs.~\ref{W-G-M}-\ref{W-G-h} and \ref{norm-m-h}
we find
\begin{equation}
    M^\delta/h =\theta^\delta \frac{h(1)}{h(\theta)} \; .
\label{MEoS}
\end{equation}

Using  Eqs.~\ref{W-G-fg} and \ref{W-G-z}
we also obtain $f'_G(z)$ as,
\begin{eqnarray}
    f'_G(z)&\equiv& \frac{{\rm d}f_G(\theta(z))}{{\rm d} z}=
    \frac{{\rm d}f_G}{{\rm d} \theta} \Big/
    \frac{{\rm d} z}{{\rm d}\theta}
    \label{fGpr}\\
    &=& 
    -\frac{\left(\theta_0^2-1\right) 
    \left(\beta\delta  h(\theta)-\beta\theta h'(\theta)\right)}{\theta_0^{1/\beta } (2 \beta  \delta 
    \theta h(\theta)-\left(\theta^2-1\right) h'(\theta))} 
    \left(\frac{h(\theta)}{h(1)}\right)^{\frac{1-\beta }{\beta  \delta }}
    \nonumber \; .
    \label{fGptheta}
\end{eqnarray}
With this the scaling function of the magnetic susceptibility, defined in Eq.~\ref{fchi}, is given by,
\begin{equation}
    f_\chi(z) = 
    \frac{\left((2 \beta-1) \theta^2+1\right) h(1)
    }{2 \beta \delta \theta h(\theta)-\left(\theta^2-1\right)
    h'(\theta)}
    \left(\frac{h(\theta)}{h(1)}\right)^{1-1/\delta} \; .
\label{fchitheta}
\end{equation}

\subsection{Analytic continuation of the scaling functions}

For real values of $z$ the scaling functions $f_\chi(z)$ and $f'_G(z)$
are finite and vanish in the limit $z\rightarrow \pm \infty$. In the 
Schofield parameterization they thus 
vanish at 
$\theta=0$ and $\theta=\theta_0$.
For complex values of $\theta$
they, however, develop singularities, which are end-points
of cuts in the complex plane. The
locations of these edge
singularities are obtained as solutions of 
\begin{eqnarray}
   0 &=& \frac{{\rm d}z(\theta)}{{\rm d} \theta} \; \nonumber \\
    \Leftrightarrow \;\;  0 &=& 2 \beta \delta \theta h(\theta)-\left(\theta^2-1\right)
    h'(\theta) \, \label{poles} \label{defzeroes}
    \label{fgzsing}
\end{eqnarray}
Well-known singularities in the
complex $z$-plane refer to the
Lee-Yang edge singularities \cite{PhysRev.87.404,PhysRev.87.410}
and the Langer edge singularity \cite{Langer:1967ax,D-S-Gaunt_1970},
\begin{eqnarray}
z_{LY} &=& |z_{LY}| {\rm e}^{i\phi_{LY}} \; ,\;
\phi_{LY} = \pi/2\beta\delta
\; , \\
z_{Lan} &=& |z_{Lan}| {\rm e}^{i\phi_{Lan}} \; ,\;
\phi_{Lan} =\pi (1-1/\beta\delta) \; ,
\end{eqnarray}
where the phases $\phi_{LY}$ and $\phi_{Lan}$ give the orientation
of cuts in the complex $z$-plane.

We want to extract here information on the location
of these singularities using an analytic 
continuation of the Schofield parameterization with input for the
functions $h(\theta)$ that has
been obtained for different
universality classes and real values
of $\theta$. 
As the mapping $z \Leftrightarrow \theta$ is in general not unique,
we will determine the region in the complex 
$\theta$-plane that provides a unique mapping
between $z$ and $\theta$ and is connected to the real interval 
$0\le \theta \le \theta_0$ onto which the
real $z$-axis is mapped. We do so by following
lines of complex $z$ values with constant phase,
\begin{equation}
    z(\theta) \equiv |z| e^{i\phi} \;\; , \;\; -\pi \le \phi < \pi \;\; .
    \label{z-mapping}
\end{equation}
All these lines start at $\theta=1$, which corresponds to $|z|=0$, and will end 
in points $\theta_n$ that correspond to zeroes
of the function $h(\theta)$.
In particular, we note that $h(\theta)$ may
change sign at $\theta_0$ and become
negative for real $\theta$ in an
interval $\theta_0<\theta<\theta_1$, 
with $\theta_1$ denoting a possibly existing 
second real zero of the function $h(\theta)$. 
In this interval one finds from Eq.~\ref{W-G-z} that $z$
becomes complex with a phase $\textrm{Arg}(z) = \pm \pi(1-1/\beta\delta)$.  
On the other hand, points on the imaginary
$\theta$-axis  correspond to complex $z$
values with phase  
$\textrm{Arg}(z) =\pm\pi/2\beta\delta$.
Certain $\theta$-intervals on the real and imaginary $\theta$-axis thus have phases corresponding
to those of the Langer and Lee-Yang cuts,
respectively. However, we note that
Eq.~\ref{z-mapping} implies that also in 
the complex $\theta$-plane lines exist
on which the corresponding $z$-values have these phase values. 
We will discuss 
implications for the location of edge singularities in the complex $\theta$-plane
in the following sections.

\section{Analytic continuation of the (generalized) linear parametric model}
\label{sec:LPM}
Approximations for the function $h(\theta)$,
which is an odd function of $\theta$ \cite{An:2017brc}, have
been derived as polynomials in $\theta$ \cite{Guida:1996ep,Campostrini:2002ky}.
To order $\theta^3$ this function depends
on a single parameter ($\theta_0$), which
gives the only non-trivial zero of $h(\theta)$. 
Having $\theta_0>1$ ensures that the real $z$-axis
can be mapped onto the interval 
$0\le \theta\le \theta_0$. This
is the so-called linear parametric model 
(LPM) \cite{PhysRevLett.22.606}. We generalize 
this ansatz here to allow $\theta_0$ to be a 
$g$-fold zero.
In the following we will discuss the two cases, 
\begin{equation}
    h(\theta)= \theta (1-(\theta/\theta_0)^2)^g \;\; , \;\; g=1,\ 2 \; ,
    \label{LPM-h}
\end{equation}
which arise as exact results for the function
$h(\theta)$ in mean-field calculations ($g=1$) and
in the $N= \infty$ limit of 3-$d$, $O(N)$
models ($g=2$). Moreover, the LPM ansatz with
$(g=1)$ also gives the exact result for
$h(\theta)$ in the 3-$d$, $Z(2)$ universality class
when calculated to ${\cal O}(\epsilon^2)$ in a 
systematic $\epsilon$-expansion \cite{PhysRevLett.29.591,Wallace:1974}.
We will discuss these cases in the 
following two subsections.

\subsection{\boldmath Mean field approximation
and large-\texorpdfstring{$N$}{N} limit of \texorpdfstring{$O(N)$}{O(N)} models}

In MFA as well as in the
$(N= \infty)$ limit of models in the
3-$d$, $O(N)$ universality class the scaling 
function $f_G(z)$ obeys a simple relation \cite{Almasi:2016gcn},
\begin{equation}
    f_G (z + f_G^2)^g = 1 \; ,
    \label{fGmfg}
\end{equation}
with $g=1$ in mean-field models and $g=2$ in the $(N= \infty)$
limit. This equation can easily be
solved for $g=1$, 
\begin{eqnarray}
f_G(z)&=& \frac{2^{1/3} \left(9+\sqrt{12 z^3+81}\right)^{2/3}-2\cdot 3^{1/3} z
    }{6^{2/3} \left(9+\sqrt{12 z^3+81}\right)^{1/3}}\ ,
\label{fGzmf}
\end{eqnarray}
from which one derives the related susceptibility scaling functions $f'_G(z)$ 
and $f_\chi(z)$ introduced in 
Eqs.~\ref{fGpr} and \ref{fchitheta}.
This, however, is not at all straightforward 
to be done for $g=2$. In that case the LPM
approximation of the magnetic equation of
state provides an easy to handle parameterization of $f_G(z)$.

Making use of Eq.~\ref{fGmfg}
and critical exponents in the mean-field  and ($N=\infty$) universality classes,
\begin{eqnarray}
\beta= \frac{1}{2}\;\; , \;\; 
\delta=
\begin{cases}
3 \;  , {\rm mean}-{\rm field}
\\
 5\;  , N=\infty
 \end{cases} \;\; ,
\end{eqnarray}
one can determine the function $h(\theta)$ entering the Schofield
parameterization of the scaling function $f_G(\theta(z))$.
We solve Eq.~\ref{W-G-fg} for $h(\theta)$
and insert this in Eq.\ref{W-G-z}. Inserting this result for $z(\theta,f_G)$ in Eq.~\ref{fGmfg}
yields
\begin{eqnarray}
f_G &=&\left( 
\left(\frac{\theta_0}{\theta} \right)^{2} \frac{1}{\theta_0^2-1}
\left(1-\left( \frac{\theta}{\theta_0}\right)^2\right)\right)^{-g/\delta} \; .
\label{fGg}
\end{eqnarray}
Here we kept $g$ and $\delta$ for
clarity. However, in deriving Eq.~\ref{fGg} we already made use of the relation $\delta=1+2 g$.
Inserting this result for $f_G$ on
the left side of Eq.~\ref{W-G-fg}
we can solve for $h(\theta)/h(1)$, 
\begin{equation}
 \frac{h(\theta)}{h(1)}= \theta
 \left(\frac{\theta_0^2}{\theta_0^2-1}\right)^{g}
\left(1-\left( \frac{\theta}{\theta_0}\right)^2\right)^{g} \; ,
\label{h-mf-ON1}
\end{equation}
or equivalently,
\begin{equation}
    h(\theta)= \theta \left(1-\left( \frac{\theta}{\theta_0}\right)^2\right)^{g} \: .
    \label{h-mf-ON}
\end{equation}
This is the linear parametric model for
the parameterization of the magnetic equation of state in the MFA ($g=1)$ and its generalization to the $N\rightarrow\infty$ limit ($g=2$).
We may use this form of $h(\theta)$ 
to analyze the analytic structure of
scaling functions in MFA and the large-$N$ limit,
using the Schofield parameterization.
In particular, we can determine the 
location of Lee-Yang edge singularities.

\begin{figure*}[t]
        \includegraphics[width=0.45\linewidth]{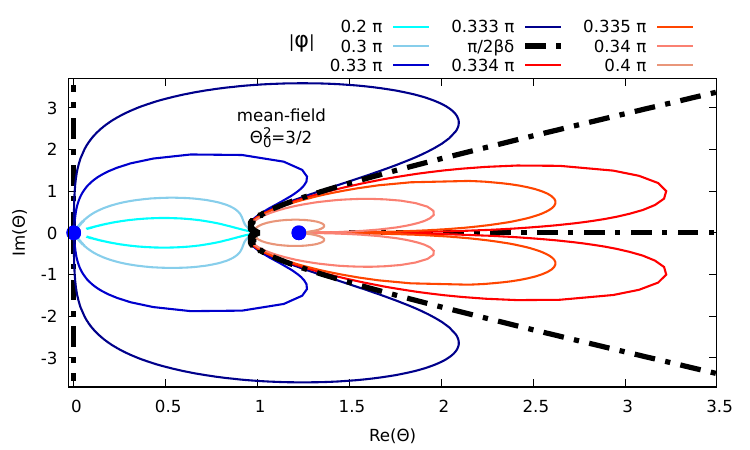}
        \includegraphics[width=0.45\linewidth]{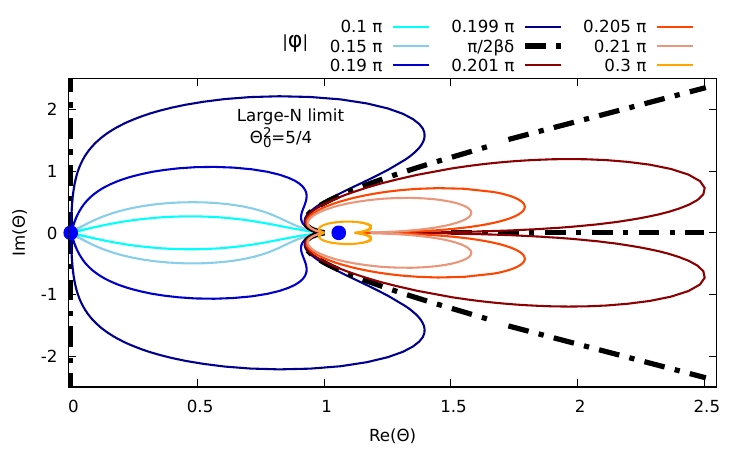}
    \caption{Mapping of the complex $z$-plane to the complex $\theta$-plane. Shown are lines
    of $z$ with constant phase $\phi$ as introduced in Eq.~\ref{z-mapping}, $z=|z|{\rm e}^{i \phi}$.
    The left hand figure is for the MFA and the right hand side shows results for the $N = \infty$ limit.
    Dash-dotted lines correspond to $z$-values having phase $\phi=\pi/2\beta\delta=\pi/\delta$. Blue dots show points at which $h(\theta)=0$.}
    \label{fig:map}
\end{figure*}

\subsubsection{Lee-Yang edge singularities}

Inserting in Eq.~\ref{defzeroes}  $h(\theta)$ taken from Eq.~\ref{h-mf-ON}  
allows to determine the location of singularities in the scaling functions in the complex $\theta$-plane.
For the generalized LPM this gives,
\begin{eqnarray}
0&=&\theta_0^2 - \left(-2 \beta  \delta \theta_0^2+
2 g+\theta_0^2+1\right) \theta ^2
    \nonumber \\
    &&+ (-2 \beta  \delta +2 g+1)\theta^4 \; .
    \label{LPMpoles}
\end{eqnarray}
In the case of the MFA and in the large-$N$
limit this quartic equation in $\theta$ reduces to a quadratic 
equation, and one thus obtains 
only one pair of singularities located at
\begin{equation}
    \tilde{\theta}_{\pm} = \pm \; \frac{\theta_0}{\sqrt{\delta- (\delta-1)\theta_0^2}}\: .
    \label{theta0mfN}
\end{equation}
Obviously,  in the complex $\theta$-plane
the location of singularities in the
susceptibilities depends
on the value of $\theta_0$.
In particular, the singular points are shifted to infinity for
\begin{equation}
\theta^2_{0,\infty}=\frac{\delta}{\delta -1} \; ,
\label{theta02}
\end{equation}
while they are real for $\theta_0<\theta_{0,\infty}$ and
purely imaginary for $\theta_0>\theta_{0,\infty}$\ .
Nonetheless, one easily verifies that
    \begin{eqnarray}
z(\theta)&=&     \frac{\left(1-\theta ^2\right) \theta_0^2}{\theta_0^2-1}  \left(\frac{\theta_0^2-\theta
    ^2}{\theta_0^2-1}\right)^{-(\delta-1) / \delta
    } \theta ^{-2 /\delta }
    \; 
    \label{z35}
\end{eqnarray}
is independent of $\theta_0$ for $\theta=\tilde{\theta}_{\pm}$. Using 
$\tilde{\theta}_{\pm}$ one finds the
universal location of the Lee-Yang edge singularity in the complex $z$-plane,
\begin{eqnarray}
    z_{LY}\equiv z(\tilde{\theta}_{\pm}) &=& \frac{\delta}{\delta-1} (1-\delta)^{1/\delta} \nonumber \\
    &=&
    \begin{cases}
         3\cdot 2^{-2/3}\ {\rm e}^{\pm i\pi  /3}\; ,\;\; {\rm mean-field} \\
    5\cdot 2^{-8/5}\ {\rm e}^{\pm i\pi  /5}\; ,\;\; N= \infty 
    \end{cases}
        \nonumber \\
\end{eqnarray}
We note that in MFA as well as in the large-$N$ limit the mapping $z\Leftrightarrow \theta$ is unique for all $z$ and maps the complex $z$-plane
to the complex $\theta$ half-plane with
$Re(\theta)\ge 0$. For the specific choice $\theta_0\equiv \pm \theta_{0,\infty}$
we show this 
mapping for both cases in Fig.~\ref{fig:map}.
The lines $z=|z|\exp(\pm i \pi/2\beta\delta)$ approach the 
Lee-Yang edge singularity at $|\theta|=\infty$ and then bifurcate.
In both cases the two branches then approach one of the two points at 
which $h(\theta)$ vanishes, {\it i.e.}
they either approach $\theta=0$ or 
$\theta_0$. These points get approached 
by the two branch cut lines in the complex $z$-plane that correspond either to the imaginary or real axis of the $\theta$-plane.

\subsection{\boldmath LPM for scaling 
functions in the  3-\texorpdfstring{$d$}{d}, \texorpdfstring{$Z(2)$}{Z(2)} 
universality class} 
\label{sec:Z2eps}

Scaling functions in the 3-$d$, $Z(2)$ universality class have been
determined using analytic as well
as numerical approaches. In particular, using the $\epsilon$-expansion, it
has been pointed out that 
to ${\cal O}(\epsilon^2)$ the LPM
approximation for the function $h(\theta)$ remains exact \cite{PhysRevLett.29.591,Wallace:1974}. 
One thus obtains $h(\theta)$ as given in
Eq.~\ref{h-mf-ON} with $g=1$ and
\begin{equation}
    \theta_0^2 =\frac{3}{2} (1-\epsilon^2/12) +{\cal O}(\epsilon^3)\; .
\end{equation}
New features arise from the fact 
the product of critical exponents,
$\beta \delta$ now deviates from $3/2$.
The coefficient of the ${\cal O}(\theta^4)$ term, appearing in Eq.~\ref{LPMpoles}, thus no longer vanishes.

Using also results for the critical exponents
to ${\cal O}(\epsilon^2)$,
\begin{eqnarray}
    \beta&=&\frac{1}{2}-\frac{1}{6}\epsilon + \frac{1}{162}\epsilon^2+{\cal O}(\epsilon^3) \; , \nonumber \\
    \delta&=& 3+\epsilon+\frac{25}{54}\epsilon^2+{\cal O}(\epsilon^3)\; \; , \\
        \beta \delta &=&
    \frac{3}{2} +\frac{1}{12} \epsilon^2 +{\cal O}(\epsilon^3)\; ,
\end{eqnarray}
we may again determine the location of singularities of the susceptibilities using Eq.~\ref{LPMpoles}. We now 
obtain two pairs of solutions for the
location of singular points in the
complex $\theta$ plane,
\begin{eqnarray}
\tilde{\theta}^2_{p,m} &=& 
\frac{A \pm \sqrt{A^2-4 B
    \theta_0^2}}{2 B} \; ,\;  
    \label{LYzeroesAB}
    \end{eqnarray}
    with 
    \begin{eqnarray}
     A&=& 1+2g +\theta_0^2-2 \beta\delta \theta_0^2\; ,
\nonumber \\
     B&=& 1+2 g - 2 \beta\delta\; .
\end{eqnarray}
Obviously, in the 3-$d$, $Z(2)$ universality
class\footnote{For the $O(N)$, $N>1$
universality classes one finds $B>0$ when $g=2$.} ($g=1$) and with the critical exponents given
in Table~\ref{tab:parameter}, one has $B<0$.

\begin{table*}[t]
\centering
\begin{tabular}{|c|c||c|c|c|c||c|} \hline
 &~M-F~&~${\cal O}(\epsilon^2), Z(2)$~& $Z(2)$ & $O(2)$ & $O(4)$&O($\infty$) \\ \hline
  $\beta$ &1/2&0.3395& 0.32643(7) & 0.34864(7) & 0.380(2)&1/2 \\
$\delta$ &3&4.4630& 4.78982(85) & 4.7798(5) & 4.824(9) &5\\
$\beta\delta$ &1.5&1.5833& 1.56354(8) & 1.6664(5) & 1.833(13) &2.5\\
\hline
\end{tabular}
\caption{Critical exponents in the
$3$-$d$, $Z(2)$, $O(2)$ and $O(4)$ universality classes.
$Z(2)$ critical exponents are taken from  
\cite{Hasenbusch:2010hkh,El-Showk:2014dwa}
and the $O(2)$ values are taken from \cite{Hasenbusch:2019jkj}. Exponents 
used for the $O(4)$ case are taken from \cite{Engels:2003nq}.
Also shown are results obtained
in 3-$d$ mean-field theory,
results from ${\cal O}(\epsilon^2)$ expansion for the 3-$d$, $Z(2)$ theory, and the $N= \infty$ limit.
}
\label{tab:parameter}
\end{table*}
The square root appearing in Eq.~\ref{LYzeroesAB} thus is positive and larger than $|A|$. Therefore, irrespective
of the value of $\theta_0$ one always finds
that $\tilde{\theta}_p^2$ and $\tilde{\theta}_m^2$
are real, but have opposite sign.
This gives rise to a pair of real as well as a pair of purely
imaginary roots, 
\begin{eqnarray}
\theta_{Lan,\pm}\equiv \pm \sqrt{\tilde{\theta}_p^2} &=& \pm \sqrt{
\frac{A + \sqrt{A^2-4 B
    \theta_0^2}}{2 B}} \; ,
    \nonumber \\
    \theta_{LY,\pm}\equiv\pm \sqrt{\tilde{\theta}_m^2} &=& \pm \sqrt{
\frac{A - \sqrt{A^2-4 B
    \theta_0^2}}{2 B}} \; .
    \label{LYzeroesAB-2}
    \end{eqnarray}
Using the ${\cal O}(\epsilon^2)$
results for $\theta_0$ and the 
critical exponents $\beta,\ \delta$
one finds for the location of these singular points
in the
$\theta$-plane\footnote{Note that $\theta_{Lan,-}$ lies outside the region used for the unique mapping $z\Leftrightarrow \theta$. }
\begin{eqnarray}
\theta_{Lan,+} &=& 2.03038 \; ,
\nonumber \\
\theta_{LY,\pm} &=& \pm 3.31198\ i\; .
\end{eqnarray}
As discussed above singular points in the $z$-plane, corresponding to purely real or imaginary $\theta$ values, have phases $\pi (1-1/\beta\delta)$
and $\pi/(2\beta\delta)$, respectively. These are the 
expected phases of cuts emerging from the Langer and Lee-Yang edge singularities.

\begin{figure}[t]
        \includegraphics[width=0.9\linewidth]{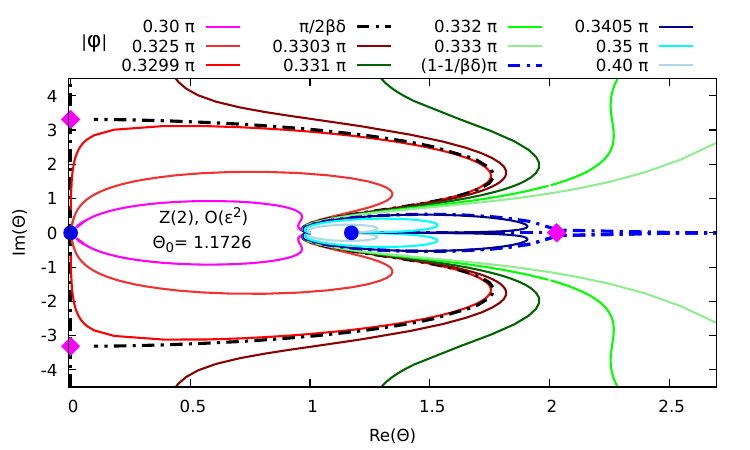}
    \caption{Contours in the complex $\theta$-plane defined by a fixed phase of $z$, $z=|z|\exp(i \phi)$ obtained by using parameters calculated
    in an $\epsilon$-expansion to ${\cal O}(\epsilon^2)$ \cite{Wallace:1974,Guida:1996ep}.
    The dash-dotted black lines corresponds to $z$-values with phase  $\phi=\pm \pi/2\beta\delta$, as
    expected for a Lee-Yang edge
    singularity
    and the dash-dotted blue line corresponds to a cut with phase as expected for the Langer cut, $\phi=\pi (1-1/\beta\delta)$.
    Blue dots correspond 
to points at which $h(\theta)=0$ and magenta diamonds correspond to $z'(\theta)=0$. 
    }
    \label{fig:Z2eps}
\end{figure}

We get for the singular points
\begin{eqnarray}
z_{Lan} &=& 2.24047 \ {\rm e}^{i\pi (1-1/\beta\delta)}\;\; ,
\nonumber \\
z_{LY,\pm} &=&  2.30674\ {\rm e}^{\pm i\pi/2\beta\delta}\;\; .
\end{eqnarray}
The corresponding mapping of the
complex $z$-plane to the $\theta$-plane is shown in
Fig.~\ref{fig:Z2eps}. As can be seen, just like in the MFA, the complex $z$-plane is 
mapped onto the entire complex
$\theta$ half-plane with ${\rm Re}( \theta)\ge 0$. 
We will discuss in the next section that this
becomes quite different when using approximations
for $h(\theta)$ that go beyond the LPM
approximation.

\section{Beyond the LPM approximation}
\label{sec:beyond}

We now consider the Schofield 
parameterization for scaling functions 
in the $Z(2)$ and $O(N)$ universality classes making use of the known critical
exponents in these universality classes (see Table~\ref{tab:parameter}) and
all parameters that have currently been determined for the approximation of the function $h(\theta)$. 
We parameterize corrections to 
the generalized LPM approximation as
an even polynomial in $\theta$ \cite{Campostrini:2000iw}
\begin{equation}
    h(\theta) = \theta \left( 1-\left(\frac{\theta}{\theta_0}\right)^2\right)^g 
    \left( 1+ c_2 \theta^2 +c_4 \theta^4
    +{\cal O}(\theta^6)\right) \; ,
    \label{hc2c4}
\end{equation}
with $g=1,\ 2$ for the $Z(2)$ and $O(N)$ universality classes, respectively.

In the case of the $Z(2)$ universality class we will also use the commonly used 
polynomial representation, where the first non-trivial
real zero of $h(\theta)$ is not factored 
out explicitly, {\it i.e} we use
\cite{Guida:1996ep,Zinn-Justin:1999opn}
\begin{equation}
    h(\theta)= \theta \left( 1+
    h_3 \theta^2+h_5 \theta^4+h_7 \theta^6 +{\cal O}(\theta^8)
\right) \; .    
\label{hth}
\end{equation}
Obviously the parameters $(\theta_0, c_2, c_4, ...)$ can be expressed in terms 
of $(h_3, h_5, h_7, ...)$. 
In particular, to this order one has $h_7 =- c_4/\theta_0^2$. We give further details on the
relation between $(h_3, h_5, h_7)$ and $(\theta_0, c_2, c_4)$
in Appendix~\ref{app:Z2}.

In the $Z(2)$ universality class
the coefficients $h_3,\ h_5$ and $h_7$
have been calculated \cite{Guida:1996ep,Guida:1998bx}.
However, usually only the coefficients $h_3$ and $h_5$,
corresponding to non-vanishing
$\theta_0$ and $c_2$, are quoted as
the coefficients $h_7$, $c_4$ are small and
vanish within current errors.
For the case of the $O(N)$ universality class ($N=2, 4$)  a first determination
of $c_4$, based on fits to Monte Carlo simulation
data for $O(2)$ and $O(4)$ scaling functions
\cite{Engels:2011km},
has been presented recently \cite{Karsch:2023pga} and also results,
based on a 3-$d$, perturbative calculation
have been obtained for the $O(2)$ universality class \cite{Campostrini:2000si}. We note that
results obtained for $\theta_0$ and $c_2$ obtained
in this calculation are in good agreement with 
the Monte Carlo results when setting $c_4=0$.
However, values for $c_4$ obtained in the analytic
calculations differ in sign from the current 
Monte Carlo results. Like in the $Z(2)$ case 
we thus expect that current results on $c_4$ 
still have large uncertainties.

In order to be able to discuss the influence of a non-vanishing next to 
leading order correction ($c_4\ne 0$) on the structure of the Schofield parameterization
of scaling functions we also calculated $c_4$ for the $Z(2)$ 
case using results obtained already in \cite{Guida:1996ep,Guida:1998bx}. 
Although $c_4$ vanishes within 
errors, including it in our analysis
allows to discuss the influence of
such a correction on the determination
of singularities of scaling functions
in the complex $z$-plane. Some details
on the calculation of $h_7$ and as 
such $c_4$ are 
given\footnote{This also allows to correct the expression given for 
$h_7$ in \cite{Guida:1996ep}, where 
a sign factor for one of the terms contributing to $h_7$
does not show up.} in Appendix~\ref{app:Z2}.
This results in the expansion
coefficients ($h_3, h_5, h_7$) in the
$Z(2)$ universality class, 

\begin{eqnarray}
    h_3&=&-0.7595(18)\; ,\; h_5= 0.00813(68)\; ,\; \nonumber \\ h_7&=&0.00045(127)\; .
    \label{h3h5h7}
\end{eqnarray}
Here we also made use of recent updates
on critical exponents in the 
$Z(2)$ universality class
\cite{Hasenbusch:2010hkh,El-Showk:2014dwa}. The effect mainly is to reduce errors on $h_3$ and $h_5$.
The parameter sets 
$(\theta_0, c_2, c_4)$, 
used in our calculations for the
determination of edge singularities 
in the  $Z(2),\ O(2)$ and $O(4)$ universality classes, are given in
Table~\ref{tab:Schofield}.

\begin{table}[t]
\centering
\begin{tabular}{|c|c|c|c|} \hline
 & $Z(2)$ \cite{Guida:1998bx} & $O(2)$ \cite{Karsch:2023pga} & $O(4)$ \cite{Karsch:2023pga} \\ \hline
  $\theta_0$& 1.1564(40) & 1.610(14)& 1.359(10) \\
$c_2$  &-0.0117(33)& 0.162(20) & 0.306(34)  \\
$c_4$ &~-0.0006(17)~ & ~-0.0226(18)~ & ~-0.00338(25)~
\\
\hline
  $\theta_1$ & 9.2(1.3)  & 1.993(86) $i$&1.777(106) $i$ \\
    $\theta_2$ & --$^\dagger$ & 3.34(21)&9.68(92) \\
\hline
\end{tabular}
\caption{Parameters entering the definition of $h(\theta)$ in the
$3$-$d$, $Z(2)$, $O(2)$ and $O(4)$ universality classes.
$Z(2)$ values are taken from  \cite{Guida:1998bx} and 
the $O(2)$ and $O(4)$ values are taken from \cite{Karsch:2023pga}. 
\\
$^\dagger$For $Z(2)$ we only quote a result for the second zero
of $h(\theta)$ obtained with $c_4=0$ as
$h_7$ and as such $c_4$ vanish within errors.
}
\label{tab:Schofield}
\end{table}

Aside from the zero $\theta_0$, which is factored out explicitly
in the representation of $h(\theta)$ given in Eq.~\ref{hc2c4}, the additional 
polynomial factor up to ${\cal O}(\theta^4)$ gives rise to
two additional zeroes in terms of $\theta^2$,
\begin{eqnarray}
    h(\theta) &=& \theta \left( 1-\left(\frac{\theta}{\theta_0}\right)^2\right)^g
    \nonumber \\
    &&\left( 1-\left(\frac{\theta}{\theta_1}\right)^2\right)
    \left( 1-\left(\frac{\theta}{\theta_2}\right)^2\right) \; .
    \label{hth1th2}
\end{eqnarray}
Results for the
zeroes, $\theta_1$ and $\theta_2$,
are also given in Table~\ref{tab:Schofield}.

The main difference in the singular structure of
the $Z(2)$ and $O(N)$ scaling functions in the complex $z$-plane arises from the
first correction to the generalized 
LPM approximation, {\it i.e.} from $c_2$
being non-zero and having different signs
in both cases. In the $Z(2)$ universality
class $c_2<0$, while $c_2>0$ in the $O(N)$ universality classes. 
For $c_4=0$ it is straightforward to
see that as a consequence the new zero, $\theta_1$, is
real in the case of $Z(2)$
and imaginary for the $O(N)$ cases.
This in turn results in the
presence of the
Langer cut \cite{Langer:1967ax} in 
the $Z(2)$ universality
class and its absence in the 
$O(N)$ universality classes:

\noindent 
As discussed in Sec.\ref{sec:Schofield},
points at which susceptibility scaling functions become
singular, are obtained as 
zeroes of $z'(\theta)$. The condition
$z'(\theta)=0$ led to Eq.~\ref{poles}, 
which we may rewrite as
\begin{equation}
z'(\theta)=0\;\; \Leftrightarrow \;\;
    \frac{h'(\theta)}{h(\theta)} = 2\beta\delta\frac{\theta}{\theta^2 -1} \; .
    \label{zerocond}
\end{equation}
This relation makes it clear that 
\begin{itemize}
\item
a singularity in the scaling functions 
exists in the interval $0<|\theta|<|\theta_1|$ on the imaginary $\theta$-axis, if $h(\theta)$ has a 
zero, $\theta_1$, on the imaginary $\theta$-axis,
\item
a singularity in the scaling 
functions is located on the real 
$\theta$-axis, if $h(\theta)$ has 
a second real zero, $\theta_1>\theta_0>1$, on the real axis.
\end{itemize}
In the $Z(2)$ case $h_5>0$ and
it follows from Eq.~\ref{Ac2} that $c_2<0$ as $c_4$ is at least an order of magnitude smaller
and vanishes within errors. One thus finds a singularity on the 
real $\theta$-axis. 
This singularity is located at a 
complex valued $z\equiv z_{Lan}$ with a phase
$\phi_{Lan}=\pi (1-1/\beta\delta)$. It is the Langer edge singularity.

In the $O(N)$ cases  $c_2>0$.
Consequently one finds a singularity on the imaginary $\theta$-axis. Actually, it is present 
for all $c_4\le 0$. This singularity is
located at a complex valued $z\equiv z_{LY,b}$, with a 
phase $\phi_{LY}=\pi/2\beta\delta$, which is 
the phase expected for a Lee-Yang edge singularity. 
Apparently this singularity in the $O(N)$ universality class
is the counter part to the Langer edge singularity
in the $Z(2)$ universality class. It
is located on the Lee-Yang cut, 
but it is not located at its edge.
The Lee-Yang edge singularity arises
from the second pair of singular points
that is present for $c_2\ne 0$ and lies
in the complex $\theta$- or $z$-plane, respectively.
This is the case in the $Z(2)$ as well as the $O(N)$ universality classes. 
The phase of $z$ corresponding to this
singular point is not immediately
apparent from the general structure
of $h(\theta)$. However, 
we show in the following that, within the current statistical and
truncation errors in the approximation for  $h(\theta)$,
the phase of $z$ is indeed consistent with $\phi_{LY}$. 
These singular points in the complex valued
$\theta$-plane with non-zero real and
imaginary parts of $\theta_{LY}$
define the Lee-Yang edge singularities
at $z\equiv z_{LY}$. We present
their determination in the $Z(2)$ and $O(N)$ universality classes separately
in the following two subsections.

\begin{figure}[t]
\includegraphics[width=1\linewidth]{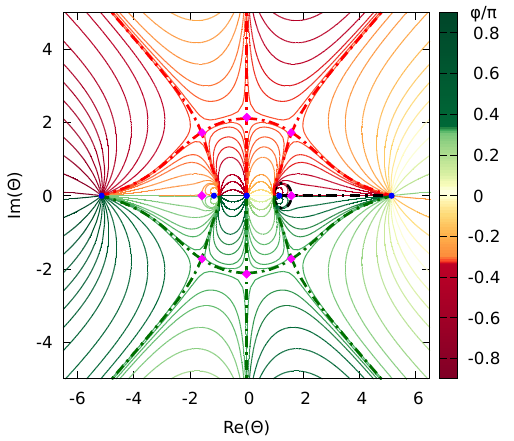}
\caption{Contours in the entire complex $\theta$-plane. Shown are lines $z=|z| {\rm e}^{i\phi}$ with constant phase $\phi\in[-\pi,\pi]$. Blue dots correspond 
to points at which $h(\theta)=0$ and magenta diamonds correspond to $z'(\theta)=0$. The dash-dotted lines correspond to the cases $\phi=\phi_{Lan}$ (black) and $\phi=\pm \phi_{LY}$ (red/green), respectively. }
\label{fig:Z2entire-plane}
\end{figure}

\begin{figure*}[t]
\includegraphics[width=0.49\linewidth]{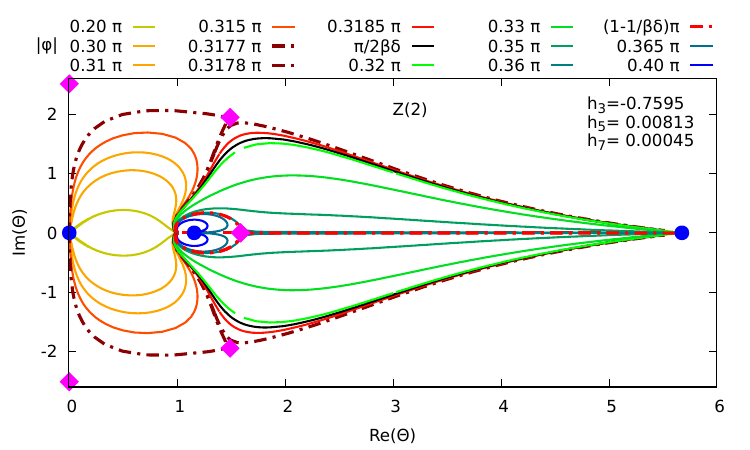}
\includegraphics[width=0.49\linewidth]{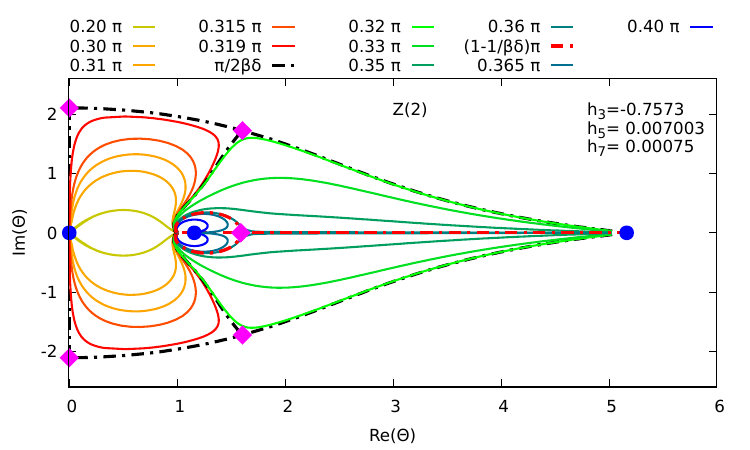}
    \caption{Contours in the complex $\theta$-plane defined by values of $z$ with constant phase $\phi$, $z=|z|\exp(i \phi)$. Shown are 
    results for the $Z(2)$ universality class using
    the function $h(\theta)$ with central values for
    $(h_3,\ h_5,\ h_7)$ given in Eq.~\ref{h3h5h7} (left)
    and tuned parameters that lead to a phase $\phi=\phi_{LY}$ (right).
    Dots show the location of zeroes of $h(\theta)$ and 
    diamonds give locations of singularities in the
    susceptibility scaling functions $f'_G(z(\theta))$ and $f_\chi(z(\theta))$.
    }
    \label{fig:Z2-mapping}
\end{figure*}

\subsection{Lee-Yang edge singularities in the \texorpdfstring{\boldmath $Z(2)$}{Z(2)} universality class}

Using the parameterization of $h(\theta)$ given in Eq.~\ref{hth} we determined the zeroes of $h(\theta)$ 
and those of $z'(\theta)$. The latter define the 
location of singular points at which the susceptibility
scaling functions $f'_G(z)$ and $f_\chi(z)$ diverge, and
the former gives points at which $z(\theta)$ diverges. In the limit $|z|\rightarrow \infty$ they
are thus attractors for lines defined by $z=|z|\exp(i \phi)$.

Contrary to the situation met in the LPM approximation 
the complex $z$-plane is no longer 
mapped onto the entire complex $\theta$ half-plane with ${\rm Re}(\theta)\ge 0$, but only to a finite 
region in that half-plane. The mapping 
$z\Leftrightarrow \theta$ is multi-valued. In Fig.~\ref{fig:Z2entire-plane} we
show the contour plot of lines, $z=|z|\exp(i \phi)$ for constant phase $\phi$ obtained in the entire $z$ plane 
for a particular set of parameters
($h_3, h_5, h_7$). The part that provides a unique mapping is shown
in Fig.~\ref{fig:Z2-mapping}.
This unique mapping, $z\Leftrightarrow \theta$,
is defined by lines 
$z=|z|\exp(i\phi)$, with $-\pi< \phi\le \pi$ emerging
from the point $\theta=1$ at which $z=0$ for all $\phi$.
These lines flow 
to $\theta=0$ or one of the two real and positive zeroes, $\theta_0$ and $\theta_1$, respectively. 

As $z'(\theta)$ is even in $\theta$ one
always finds sets of zeroes with 
positive and negative ${\rm Re}(\theta)$.
As a unique mapping of the complex
$z$-plane to the $\theta$-plane is already obtained by
keeping only zeroes with ${\rm Re}(\theta)\ge 0$, we
will quote in the following only zeroes of $h(\theta)$ and  $z'(\theta)$ with ${\rm Re}(\theta)>0$.
We also note that due to the fact that 
$h(\theta)$ is an odd function in $\theta$
$z$-values obtained from $\theta$ and its 
negative counterpart are related to each
other by
\begin{equation}
    z(-\theta)= z(\theta)\ {\rm e}^{- i s/\beta\delta} \;\; ,\;\ s=\textrm{Arg}(\theta)/|\textrm{Arg}(\theta)| .
\end{equation}

The region providing a unique mapping of
the entire $z$-plane
to the complex $\theta$-plane is bounded by lines
corresponding to $z(\theta)=|z|{\rm e}^{\pm i r\phi_{LY}}$, which bifurcate at
the complex points $\theta_{LY}$
and $\theta_{LY}^*$, respectively. 
For $r=1$ these points
correspond to the Lee-Yang edge singularities, located in the complex $z$-plane at $z_{LY}=|z_{LY}|\exp(\pm i\ r\phi_{LY})$ (Fig.~\ref{fig:Z2-mapping}~(right)). 
However, as will be discussed below, due to 
truncation errors for the function $h(\theta)$ and
statistical errors on the expansion parameters used
in the definition of $h(\theta)$ one generally only finds 
$r\simeq 1$ (Fig.~\ref{fig:Z2-mapping}~(left)).
At the bifurcation point $\pm \theta_{LY}$ the two emerging branches define branch cuts in the 
complex $z$-plane with phase $\phi_{bi}\equiv r \phi_{LY}$. The scaling function
$f_G(z(\theta))$ is discontinuous across this cut as well as across 
the Langer cut.

In Table~\ref{tab:Z2singularities} we give results
for the Langer and Lee-Yang edge singularity obtained
by (i) setting $c_2=c_4=0$ (LPM approximation), 
(ii) setting only $c_4=0$, and (iii) keeping all
three coefficients in the parameterization of
$h(\theta)$ to be non-zero, respectively. 
As can be seen, the results for the
absolute values of the edge singularities
are fairly stable. The phase of the
singularity, identified as the Lee-Yang 
edge singularity, agrees with $\phi_{LY}=\pi/2\beta\delta$ to better 
than 1\%.

\begin{table}[t]
\centering
\begin{tabular}{|c||c|c|c|} \hline
~& $|z_{Lan}|$ & $|z_{LY}|$ & $r\equiv \phi_{bi}/\phi_{LY}$\\[2pt]
\hline
$c_2=0,\ c_4=0$ & 2.5429(8) & 2.3177(40) & 1  \\[2pt]
$c_2\ne 0$,\ $c_4=0$ & 2.3693(93) & 2.4398(63) & 0.9855(23)\\[3pt]
$c_2\ne 0$,\ $c_4\ne 0$ & 2.379(36)& 2.418(55) & $0.9935^{+45}_{-191}$ \\[3pt]
\hline
Fig.4 (left) & 2.3783 & 2.4177 & 0.9935 \\
Fig.4 (right)& 2.3827 & 2.4019 &1\\
\hline
\end{tabular}
\caption{Absolute values of the Langer and Lee-Yang
edge singularities obtained with different
approximations for the function $h(\theta)$. The
last column gives results for the phase $\phi_{bi}$ of
the complex valued $z$ at the bifurcation point. This should equal 
the Lee-Yang phase $\phi_{LY}=\pi/2\beta\delta$ with
better controlled approximations for $h(\theta)$.
}
\label{tab:Z2singularities}
\end{table}
Already for $c_2=c_4=0$ one obtains 
a pair of complex (purely imaginary) valued
singular points, $\theta = \pm \theta_{LY}$, 
and a real, positive one at $\theta\equiv \theta_{Lan}$, which correspond to the
location of the Lee-Yang and Langer edge
singularities, respectively. For $c_2\ne 0$ 
the singular points $\pm \theta_{LY}$  move into the complex plane.

With $c_4\ne 0$ the location of
the value of $\theta_{LY}$
and as such $z_{LY}(\theta_{LY})$ changes
only slightly. In addition
another pair of singularities shows up at $\theta_{LY,b}$. 
Given the currently large errors 
on $c_4$ the location of this
second pair of zeroes of $z'(\theta)$ is 
not well controlled. 
For $c_4>0$ the singularity 
is located on
the real $\theta$-axis. For 
$c_4<0$ it lies on the imaginary
axis. 

While variations
of $c_4$ within its current errors
thus influences the singular structure 
in the $\theta$-plane, the location
of Lee-Yang and Langer edge singularities at $\theta_{LY}$ gets modified 
only little. The main effect of 
a variation of $c_4$ within its 
errors is to increase the current
error on the location of $z_{LY}\equiv z(\theta_{LY})$
and $z_{Lan}\equiv z(\theta_{Lan})$. 
We obtain
\begin{eqnarray}
z_{LY} &=& 2.418(55)\ {\rm e}^{\pm i r\phi_{LY}}
\; ,\; r =0.9935^{+45}_{-191}
\;\; ,
\nonumber \\
z_{Lan} &=& 2.379(36)\ {\rm e}^{i \phi_{Lan}}   \; . 
\label{Z2edges}
\end{eqnarray}
In Fig.~\ref{fig:Z2-mapping}~(left)
we show the unique region for the mapping $z\Leftrightarrow \theta$
and the location of zeroes and singular points
in the $\theta$-plane corresponding
to the central values of ($h_3, h_5, h_7$)
given in Eq.~\ref{h3h5h7} that have been used to determine the location of the 
edge singularities given in Eq.~\ref{Z2edges}. In the right hand figure we show results for a tuned set of
parameters ($h_3, h_5, h_7$), which
leads to a bifurcation point having 
the correct phase of a Lee-Yang edge 
singularity. 
Branch cuts emerge from these 
edge singularities located on 
the dashed-dotted lines, which correspond to contours having the phase $\pm \phi_{LY}$ (black)
and $\phi_{Lan}$ (red), respectively.

\begin{figure}[t]
     \includegraphics[width=0.95\linewidth]{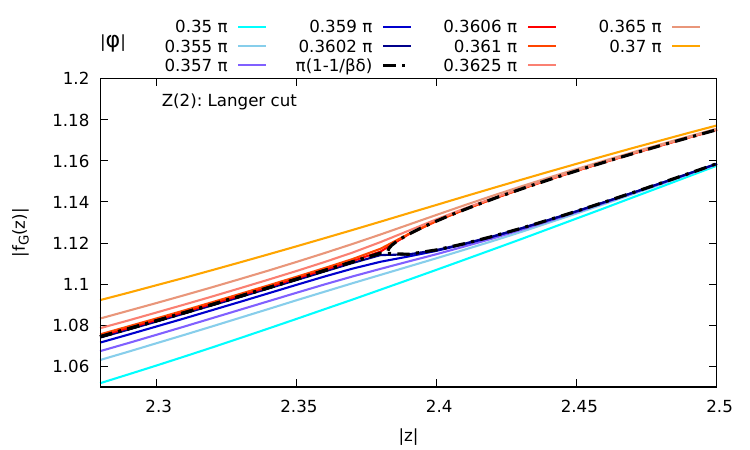} 
     \includegraphics[width=0.95\linewidth]{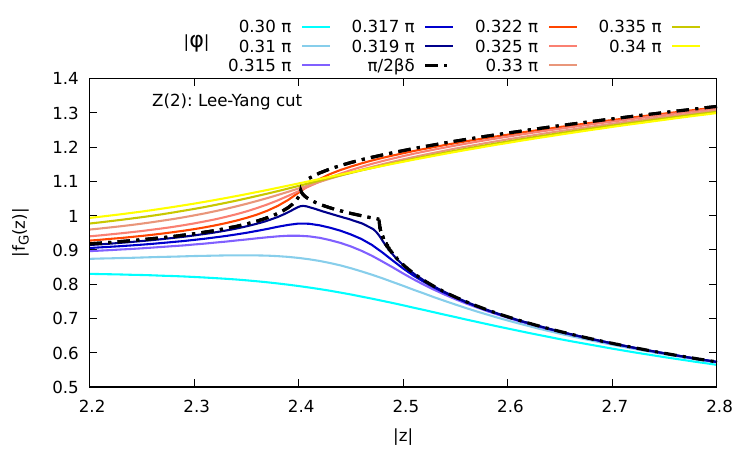}
    \caption{The absolute value of the $Z(2)$ scaling function $f_G(z)$ on the Langer (top) and Lee-Yang (bottom) cuts. 
    Shown are results for the $Z(2)$ universality class
    using $h(\theta)$ with the tuned 
    set of parameters ($h_3, h_5, h_7$)
    also used in Fig.~\ref{fig:Z2-mapping}~(right).
}
    \label{fig:branch}
\end{figure}

Starting from the edge singularities
branch cuts emerge. These lead to discontinuities in the real and imaginary part of the scaling function 
$f_G(z)$. 
In Fig.~\ref{fig:branch} we show the gaps of $|f_G(z)|$  on the Langer and Lee-Yang branch cuts, respectively. As can be 
seen the magnitude of the discontinuity 
is quite different on these cuts. It is
an order of magnitude larger on the 
Lee-Yang cut than on the Langer cut,
which may be taken as an indication for the 
Langer edge singularity being an essential singularity \cite{Langer:1967ax} which leads to only
a weak discontinuity across the cut.
This, however, is not accessible with
the truncated polynomial expansion used
for $h(\theta)$. A more complicated form
of $h(\theta)$ will be needed to reproduce the essential singularity at the Langer cut, as well as the universal
form of the Lee-Yang edge singularity, which is expected 
to be described by a  $\phi^3$-theory
\cite{PhysRevLett.40.1610,Fonseca:2001dc,An:2017brc}.

We conclude that results for the location of edge
singularities in the $Z(2)$ universality class are
quite stable and vary only little when
adjusting the phase of $z(\theta_{LY})$. Even when
moving from the LPM approximation 
to the currently known parameterization
of the function $h(\theta)$ up to ${\cal O}(\theta^7)$ the variation
of $|z_{LY}|$ is only about 5\%.

\subsection{Lee-Yang edge singularities in the \texorpdfstring{\boldmath $O(N)$}{O(N)} universality classes}

Aside from a 2-fold zero
in the generalized LPM approximation for the 
function $h(\theta)$, which also
is implemented in the general
ansatz for $h(\theta)$ \cite{Campostrini:2000si},
the most important difference in corrections to
the LPM approximation for $Z(2)$
and $O(N)$ universality classes, respectively, 
is due to the different sign of the coefficient $c_2$ in the leading order correction. 
As discussed above a consequence of this is that one 
always finds a pair of complex conjugate 
singularities in the susceptibility scaling functions
located on the imaginary $\theta$-axis
at $\pm \theta_{LY,b}$,
which corresponds to complex $z$-values with
the Lee-Yang phase $\phi_{LY}=\pi/2\beta\delta$.
Moreover, already for $c_2=c_4=0$
one obtains a pair of singularities 
located at $\theta_{LY}$ in the 
complex $\theta$-plane.
As discussed in the previous subsection for the case of Lee-Yang singularities in the $Z(2)$ universality class, also
in the $O(N)$ case
the phase $\phi_{bi}$ at this singular point,
$z(\theta_{LY})$, is not exactly at $\phi_{LY}$.
However, already for $c_2\ne 0$ the phase $\phi_{bi}$ agrees
with $\phi_{LY}$ within current statistical errors
and this remains to be the case for $c_2\ne 0$ and
$c_4\ne 0$. This behavior is found for the $O(2)$
as well as the $O(4)$ universality classes. 

\begin{figure*}[t]
     \includegraphics[width=0.59\linewidth]{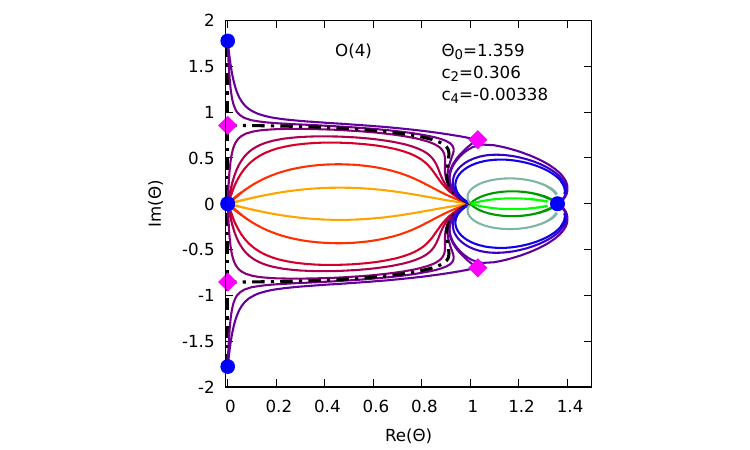}\hspace*{-4.0cm}
     \includegraphics[width=0.59\linewidth]{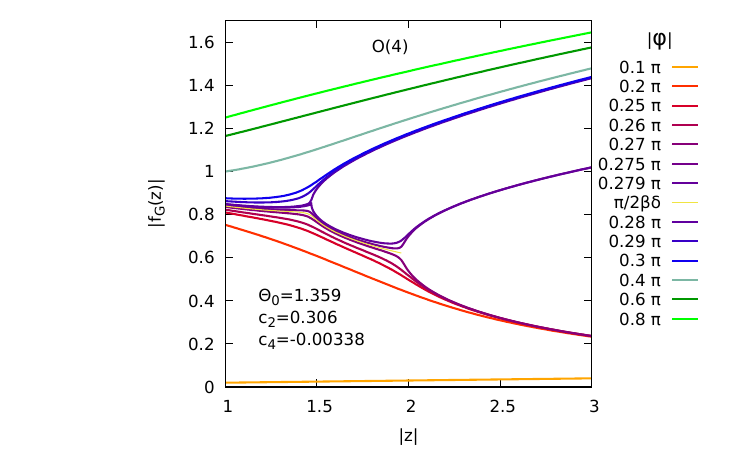}
\includegraphics[width=0.59\linewidth]{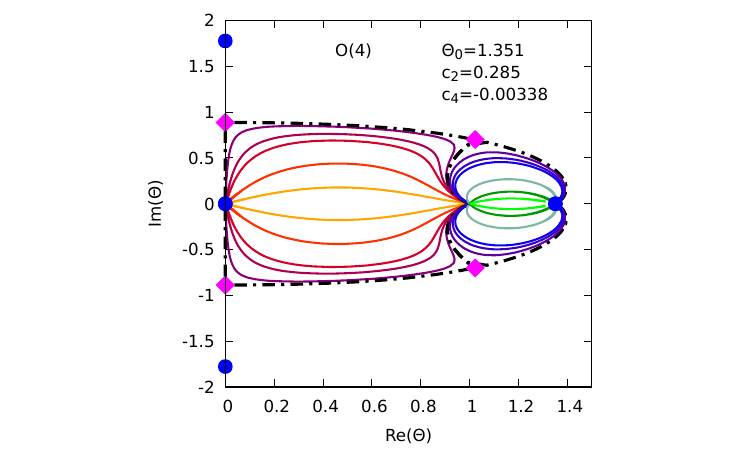} \hspace*{-4.0cm}
     \includegraphics[width=0.59\linewidth]{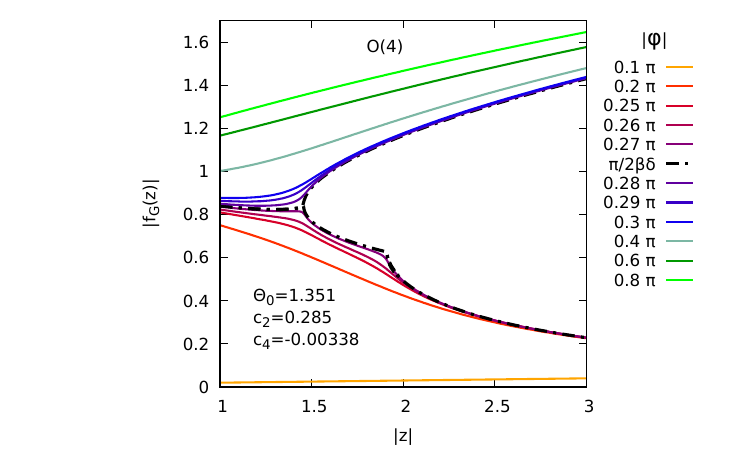}
    \caption{Contours in the complex $\theta$-plane defined by values of $z$ with constant phase $\phi$, $z=|z|\exp(i \phi)$.
    Shown are results for the $O(4)$ universality class
    using $h(\theta)$ with the central values for ($\theta_0, c_2, c_4$) given in Table~\ref{tab:Schofield}
    (top, left)
    and for a specific nearby choice of parameters (within
    the current errors) that correspond to $\phi=\phi_{LY}$
    (bottom, left). 
    The figures on the right show $|f_G(z)|$ for
    both cases. In the left hand figures dots show
    the location of zeroes of $h(\theta)$ and 
    diamonds give locations of singularities in the
    susceptibility scaling functions $f'_G(z(\theta))$ and $f_\chi(z(\theta))$.
    Similar results can be obtained in the $O(2)$ universality class.}
    \label{fig:O4-mapping}
\end{figure*}

As has been done in the previous subsection for the $Z(2)$ universality class
we also determined the absolute values of the Lee-Yang phase and the 
values  for the phase 
$\phi_{bi}=r \phi_{LY}$ 
for three different cases: (i) $c_2=c_4=0$, (ii)
$c_2=0$, $c_4\ne 0$, (iii) $c_2\ne 0$, $c_4\ne 0$.
Our results are summarized
in Table~\ref{tab:ONsingularities}.
In analogy to Fig.~\ref{fig:Z2-mapping}
we show in Fig.~\ref{fig:O4-mapping} 
contour plots in the $O(4)$ universality 
class. Fig.~\ref{fig:O4-mapping} (top, left) 
is for the set of central values
for $(\theta_0, c_2, c_4)$
corresponding to $r=1.023$ and 
the (bottom, left) figure corresponds
to a parameter set with $\phi_{bi}=\phi_{LY}$, which
lies inside the region defined by the current errors on $(\theta_0, c_2, c_4)$.

We note that the absolute value of the
Lee-Yang edge singularity changes little when using only the LPM approximation for $h(\theta)$ or
including the parameters $c_2$ and $c_4$.

\begin{table}[t]
\centering
\begin{tabular}{|c||c|c||c|c|} \hline
&\multicolumn{2}{|c|}{$O(2)$}
&\multicolumn{2}{|c|}{$O(4)$}\\
\hline
~& $|z_{LY}|$& $\phi_{bi}/\phi_{LY}$ & $|z_{LY}|$ & $\phi_{bi}/\phi_{LY}$\\[2pt]
\hline
$c_2=0,\ c_4=0$ & 1.815(27) & 0.868(7) & 1.387(15) & 0.789(9)  \\[2pt]
$c_2\ne 0$,\ $c_4=0$ & 1.999(44) & 1.037(33)& 
1.474(20) & 1.027(34)\\[3pt]
$c_2\ne 0$,\ $c_4\ne 0$ & 1.900(46)& 1.024(30) & 
1.469(20)& 1.023(34)\\[3pt]
\hline
Fig.~\ref{fig:O4-mapping}~(top) & -- & -- & 1.469& 1.023 \\
Fig.~\ref{fig:O4-mapping}~(bottom)& --& -- & 1.456 &1\\
\hline
\end{tabular}
\caption{Absolute values 
of the  Lee-Yang edge singularities in the $O(2)$ 
and $O(4)$ universality classes obtained with different approximations for the function $h(\theta)$.
}
\label{tab:ONsingularities}
\end{table}

In the right hand column of 
Fig.~\ref{fig:O4-mapping}
figure we show the absolute value of $f_G(z(\theta))$, evaluated on lines with constant phase $z=|z|{\rm e}^{i \phi}$, $\phi\simeq \phi_{bi}$, versus $|z|$.
The gap along the Lee-Yang branch cut clearly is visible. It is of similar magnitude as the discontinuity found in the $Z(2)$ universality class. 
We note that results obtained with the tuned parameter set ($\theta_0, c_2, c_4$) and the central values
determined for ($\theta_0, c_2, c_4$) 
differ little in the location of 
the edge singularity as well as the size
of the gap along the Lee-Yang branch cut.

\section{Conclusions and Outlook}
\label{sec:conclusion}

We give our final results for the
location of Lee-Yang edge singularities
of scaling functions in the 
$Z(2)$, $O(2)$ and $O(4)$ universality classes as well as the Langer edge 
singularity appearing in the
scaling functions of the 
$Z(2)$ universality class in Table~\ref{tab:singularities}.
Here we have averaged over the results
obtained when setting $c_4=0$ or using
$c_4\ne 0$ as discussed in the previous section. The difference between these 
two results is included as a systematic
error that is added to the statistical 
error in quadrature.
These results are compared with results 
obtained in FRG calculations
\cite{Connelly:2020gwa,Johnson:2022cqv},
which are given in the last row of 
Table~\ref{tab:singularities}.
As can be seen results in the $Z(2)$ and 
$O(2)$ universality classes, obtained from 
the Schofield parameterization of the 
magnetic equation of state and the FRG
calculations, agree within the quoted errors
while the result obtained in the $O(4)$
universality class differ by about 15\%. 

Also shown in the first row of Table~\ref{tab:singularities}
are results for the additional
singularity at $z_{LY,b}$ that arises
on the branch cuts. To what extent these singularities are
artifacts of our current
approximation for $h(\theta)$
remains to be analyzed by using 
higher order approximations for
$h(\theta)$.

Also shown in Table~\ref{tab:singularities}
is the absolute value,
$|z_{Lan}|$, for the location of the Langer edge singularity in.
Although its central value is smaller than that for
the location of the Lee-Yang edge singularity, it is consistent with 
$|z_{LY}|$ within errors.

We also have shown that the discontinuity of $f_G(z)$ on the 
Langer branch cut is an order of magnitude 
smaller than the discontinuity occurring on the Lee-Yang branch cut.
This may reflect the different critical
behavior expected to control the divergence of 
susceptibilities at the edge singularities. This, however, cannot be
resolved with the truncated series approximation used for the function $h(\theta)$ appearing in the Schofield 
parameterization of scaling functions. At present, the 
singular behavior at the edge singularities is identical at
the Langer and Lee-Yang edge singularity as well as in all 
universality classes, due to the polynomial ansatz used for $h(\theta)$. This is because in all cases the singularity is determined by a zero
of Eq.~\ref{fgzsing}. With $h(\theta)$ being a polynomial in $\theta$ one 
can Taylor-expand in the vicinity of any of these zeroes, 
which generates a divergence at the Lee-Yang or Langer edge
singularities being given by $1/(\theta-\theta_i)$, with $i=LY$or $Lan$.
It would be interesting to explore more
refined analytic ans\"atze for $h(\theta)$
in the future.

All data from our calculations, presented in the figures of this paper, can be found in Ref.\cite{DataSetLY}

\begin{table}[t]
\centering
\begin{tabular}{|c|c|c|c|c|} \hline
 &$Z(2)$: ${\cal O}(\epsilon^2)$ & $Z(2)$ & $O(2)$  & $O(4)$  \\ \hline
 $|z_{LY,b}|$ & -- & -- & 2.281(72) &  1.977(73)\\ \hline
$|z_{Lan}|$ & 2.240 & 2.374(36) & -- & -- \\ 
$|z_{LY}|$ & 2.307 & 2.429(56) & 1.95(7) & 1.47(3) \\
\hline
~$|z_{LY}|~[FRG]~$ & & 2.43(4) & 2.04(8) & 1.69(3) \\
\hline
\end{tabular}
\caption{Summary of results for absolute values of the location of the Lee-Yang edge singularities ($|z_{LY}|$
and the Langer edge singularity ($|z_{Lan}|$. In the first row we give 
the value for the location of a second
singularity on the the Lee-Yang branch 
cut ($|z_{LY,b}|$ (see discussion in the text). The last row gives results from
a FRG calculation \cite{Johnson:2022cqv}.
}
\label{tab:singularities}
\end{table}

\vspace{0.5cm}
\emph{Acknowledgments.---} 
This work was supported by The Deutsche Forschungsgemeinschaft (DFG, German Research Foundation) - Project number 315477589-TRR 211 and the PUNCH4NFDI consortium
supported by the Deutsche Forschungsgemeinschaft (DFG, German Research Foundation) with project number 460248186 (PUNCH4NFDI). 
\vspace{0.2cm}

\appendix

\section{\texorpdfstring{\boldmath $Z(2)$}{Z(2)} parameters}
\label{app:Z2}
We summarize here the determination 
of $h_3$, $h_5$ and $h_7$ entering 
the parameterization of the function $h(\theta)$ in the $Z(2)$ universality
class. We follow
Ref.~\cite{Guida:1998bx}.

We use the critical exponents, also used 
in that calculation
\begin{eqnarray}
\beta &=&0.3258(14)\; , \nonumber \\
\gamma &=&1.2396(13) \; .
\label{Guida}
\end{eqnarray}
In addition we also give results obtained by using
recent, more accurate results for
the critical exponents, obtained
in Monte Carlo
\cite{Hasenbusch:2010hkh} and conformal bootstrap \cite{El-Showk:2014dwa} calculations.
Both approaches yield consistent results with similarly small errors. The latter gives
$\nu=0.62999(5)$, $\eta=0.03631(3)$. Using
hyper-scaling relations we obtain 
\begin{eqnarray}
\beta &=& 0.32643(7) \; , \nonumber \\
\gamma&=& 1.23711(12) \; .
\label{bootstrap}
\end{eqnarray}
Using the hyper-scaling relation,
$\delta=(\gamma+\beta)/\beta$, we
obtain 
$\delta=4.8048(20)$ using Eq.~\ref{Guida}  and 
$\delta=4.7898(12)$
using Eq.~\ref{bootstrap}.

We compare results for $(h_3, h_5,h_7)$ obtained with
these updated critical exponents to those obtained in Ref.~\cite{Guida:1998bx} in Table~\ref{tab:h3h5h7}.

The coefficients $(h_3, h_5, h_7)$ have been determined
in \cite{Guida:1998bx} using resummed results
of a $3$-$d$ 
perturbative expansion of the 
function $F(z)$, which is defined as
derivative of the free energy with 
respect to a variable $\tilde{z}$,
\begin{equation}
    F(\tilde{z}) = \tilde{z} +\frac{1}{6} \tilde{z}^3 + F_5 \tilde{z}^5 + F_7 \tilde{z}^7 \; ,
\end{equation}
with 
\begin{equation}
F_5=0.01711(7)\;\; ,\;\; F_7 =0.00049(5) \; .
\end{equation}
The expansion parameter $\tilde{z}$
is related to the variable 
$\theta$, used in the Schofield parameterization, through,
\begin{equation}
    \tilde{z}=\rho \theta/(1-\theta^2)^\beta \; .
\end{equation}
In \cite{Guida:1998bx} the scale parameter
$\rho^2=2.8656$ is used.
We assign an error of $10^{-2}$ to it\footnote{No error on $\rho^2$ has been quoted in \cite{Guida:1998bx}. The error assigned by us, however, reproduces the error 
on $h_3$ and $h_5$ given in that reference.}.

The function $F(\tilde{z})$
is related to $h(\theta)$ through
\begin{equation}
    h(\theta)=\rho^{-1} (1-\theta^2)^{\beta\delta} F(\tilde{z}(\theta))\; .
\end{equation}
Expanding the right hand side of this
equation in terms of $\theta$ and 
using 
\begin{equation}
h(\theta)=\theta (1+ h_3 \theta^2
+ h_5 \theta^4 + h_7 \theta^6 + ...)
\; ,
\end{equation}
one arrives at relations 
for $(h_3, h_5, h_7)$ in terms of the expansion coefficients $(F_3, F_5, F_7)$,
\begin{eqnarray}
h_3 &=& \frac{1}{6} \rho ^2- \gamma \; , \\
h_5 &=& \frac{1}{2} \gamma \left(\gamma -1\right)+\frac{1}{6}\left( 2\beta-\gamma\right)\rho^2  
+\text{F}_5\ \rho ^4 
 \; , \\
h_7 &=&
 -\frac{1}{6} \gamma  \left(\gamma-1\right)\left(\gamma-2 \right) 
 \nonumber \\
 &&+ \frac{1}{12} \left(2 \beta -\gamma\right)\left( 2\beta- \gamma+1 \right)
 \rho^2
 \nonumber \\
 &&+ (4 \beta -\gamma ) \text{F}_5\   \rho^4
 +\text{F}_7\ \rho^6  \; .
\end{eqnarray}
\begin{table}[t]
    \centering
    \begin{tabular}{|c|c|c|}
    \hline
        $h_3$ & $h_5$ & $h_7$  \\
        \hline
         -0.7620(30) & 0.00818 (92)& 0.00024 (128) \\
         -0.7595(18)& 0.00813(68)& 0.00045 (127)\\
         \hline
    \end{tabular}
    \caption{Expansion coefficients of $h(\theta)$. The first row gives results obtained by using the critical exponents determined in \cite{Guida:1998bx} and the last row uses the bootstrap results from \cite{El-Showk:2014dwa}.}
    \label{tab:h3h5h7}
\end{table}

Note that the sign of the 
first term in the relation for $h_7$ is opposite 
to that quoted in  \cite{Guida:1998bx}. 
Using these relations we reproduce the expansion
coefficients $h_3,\ h_5$, given in
\cite{Guida:1998bx} and obtain results for
$h_7$ consistent with statements made in the
\cite{Guida:1998bx} about the magnitude of $h_7$. These numbers are 
given in Table~\ref{tab:h3h5h7}. We also
reproduce the errors quoted for $h_3$ and $h_5$ 
in\cite{Guida:1998bx}.

Using results for $(h_3, h_5, h_7)$ we can calculate
the coefficients $(\theta_0, c_2, c_4)$ appearing in 
the parameterization of $h(\theta)$ given in
Eq.~\ref{hc2c4}. For $c_2$ and $c_4$ one obtains,
\begin{eqnarray}
c_2 &=& - \left(h_5+h_7 \theta_0^2\right) \theta_0^2\; , \label{Ac2}\\
c_4 &=& - h_7 \theta_0^2 
\;\; ,
\label{Ac4}
\end{eqnarray}
and $\theta_0$ is obtained as the real,
positive zero of $h(\theta)$, defined in Eq.~\ref{hth},
which is closest to $\theta=1$. 

Using the parameters given in the last row of Table~\ref{tab:h3h5h7} we obtain
for the first zero of $h(\theta)$,
\begin{equation}
    \theta_0=1.1564(39)\; ,
\end{equation}
and the absolute value of the Langer edge
singularity is,
\begin{equation}
    |z_{Lan}|=2.379(36) \, .
\end{equation}
We furthermore find two singular points with a phase that is  
close to or identical to that expected for the Lee-Yang cut, $\phi_{LY}=\pi/2\beta\delta$. 
The edge singularity has an absolute value,
\begin{equation}
    |z_{LY}|=2.418^{+0.043}_{-0.068} \; ,
\end{equation}
with a phase that is consistent with $\phi_{LY}$. 
For the second singular point the absolute value is,
\begin{equation}
    |z_{LY,b}|=2.452^{+0.034}_{-0.012} \; ,
\end{equation}
and the phase equals $\phi_{LY}$.

\bibliography{LY}

\end{document}